\documentclass[twocolumn,twoside,slac_two]{revtex4}
\usepackage{graphicx}
\usepackage{fancyhdr}
\newcommand{\ms}{$M_{\odot}$}

\newcommand{\lumcgs}{ergs~s$^{-1}$}
\newcommand{\apj}{Astrophys. J.}

\begin{document}

\title{X-Ray Studies of Redbacks}

%

\author{Mallory S.E. Roberts}
\affiliation{New York University Abu Dhabi, Abu Dhabi, UAE / Eureka Scientific, Oakland, CA. USA}
\author{Maura A. McLaughlin, Peter A. Gentile}
\affiliation{West Virginia University, Morgantown WV, USA}
\author{Paul S. Ray}
\affiliation{Naval Research Laboratory, Washington DC, USA}
\author{Scott M. Ransom}
\affiliation{National Radio Astronomy Observatory, Charlottesville VA, USA}
\author{Jason W.T. Hessels}
\affiliation{ASTRON, Dwingeloo, Netherlands}
%

\begin{abstract}

We consider the X-ray properties of the redback class of eclipsing millisecond pulsars. These are transitional systems between accreting low-mass X-ray binaries and binary millisecond pulsars orbiting white dwarfs, and hence their companions are non-degenerate and nearly Roche-lobe filling. The X-ray luminosity seems to scale with the fraction of the pulsar sky subtended by the companion, suggesting the shock region is not much larger than the companion, which is supported by modeling of the orbital light curves.  The typical X-ray photon spectral index is $\sim 1$ and the typical 0.3-8~keV X-ray efficiency, assuming a shock size on the order of the companion's Roche lobe cross-section, is on the order of 10\%. We present an overview of previous investigations, and present new observations of two redbacks, a Chandra observation of PSR J1628$-$3205 and a XMM-Newton observation of PSR J2129$-$0429. The latter shows a clearly double peaked orbital light curve with variation of the non-thermal flux by a factor of $\sim 11$, with peaks around orbital phases 0.6 and 0.9. We suggest the magnetic field of the companion plays a significant role in the X-ray emission from intrabinary shocks in redbacks.

\end{abstract}

\maketitle

\thispagestyle{fancy}


\section{The Redback Population}

Millisecond pulsars are thought to be formed in binary systems where an old neutron star is spun-up via long term accretion from an evolved companion.  
In recent years, the  MSP recycling scenario has been dramatically confirmed through observations of so-called ``redback" millisecond pulsar systems \citep{rob11} which have non-degenerate companions and in some cases  transition between states with no visible radio pulsations but with optical and X-ray evidence of an accretion disk, and a state where radio pulsations are observed that regularly eclipse near superior conjunction. The first of these transition objects, PSR J1023+0038, showed optical evidence for an accretion disk in 2001 which had disappeared by 2004 \citep{tho05} . In 2007, radio pulsations were discovered \citep{arc09}, and in 2013 the MSP returned to an accreting state \citep{sta14}.

Millisecond pulsars in compact binary systems have the potential of providing unique insights into pulsar winds. The companion forces a shock to occur at a distance $d_s$ only $\sim 10^4$ times the light cylinder radius of the pulsar $R_{lc}=P_sc/2\pi$ (where $P_s$ is the spin period, and $c$ the speed of light), as compared to the more typical $d_s\sim 10^8-10^9 R_{lc}$ of the termination shock of pulsar wind nebulae around young, isolated pulsars. This means that the shock probes the wind in a region which might be significant in determining how the magnetization parameter  $\sigma$, the ratio of magnetic energy to kinetic energy, goes from a presumably high value at the light cylinder to an apparently low value at the termination shock in typical pulsar wind nebulae \citep[cf.][]{kc84}.
The basic shock emission theory for such intrabinary shocks has generally followed the outline of  \citet{at93} first developed for the original black widow system. In this model, the pulsar wind shocks with material ablated from the companion's surface, which is presumably swept back around the companion and ejected from the system. In these models, it is generally assumed that the only significant source of magnetic field is the magnetization of the wind, and that the X-ray emission is synchrotron which can be somewhat beamed either through a partially ordered magnetic field or doppler boosting.

 A {\it Chandra} observation of PSR J1023+0038 in its radio pulsar state
revealed significant orbital variability over five consecutive orbits \citep{bog11}, with a pronounced dip in the X-ray flux at superior conjunction, when the  
companion is between the pulsar and observer and the intrabinary shock produced through the interaction of stellar outflows is obscured.  The X-ray spectrum consists of a dominant non-thermal component from the shock and at least one thermal component, likely originating from heated pulsar polar caps. The eclipse depth and duration imply that the shock is localized near or at the companion surface. However, the companion only subtends $\sim 1\%$ of the pulsar's sky, so that if the wind is isotropic, only $\sim 1\%$ of the pulsar's wind is intercepted by the companion, and only $\sim7\%$ would be intercepted if the wind is confined to an equatorial sheet. \citet{bog11} inferred a high $\sigma$ from the estimated  magnetic field of $\sim 40$G required to account for the soft X-ray luminosity. 
 
An observation with \textit{NuSTAR} of PSR J1023+0038 just before it returned to the accreting state \citep{ten14} showed that the spectrum of the intrabinary shock is a very hard power law (photon index $\Gamma = 1.17$) with no apparent cutoff out to $\sim 50$keV, for a remarkable X-ray efficiency of $\sim 2\%$ of the total spin down power, or around all of the nominal spin down power in the wind that would be intercepted by the companion.  Such a hard spectrum is not easily obtained from a pulsar wind nebula shock, and such efficiency is unprecedented. This might be an indication of a significant equatorial enhancement in the wind or a significantly higher moment of inertia than the canonical $10^{45} {\rm gm\, cm^2}$, but it is still a remarkably high efficiency under any circumstances. 

Systematic studies of X-ray emission from redbacks show some commonalities.  \citet{lin14} examined the \textit{Swift} XRT data on redbacks and noted that, while in the pulsar state, their 0.5-10 keV luminosities tend to be in the  range of $L_x\sim 10^{32} {\rm erg}\,{\rm s}^{-1}$ divided into relatively high luminosity  ($L_X \gtrsim 10^{32}$\lumcgs ) and relatively low ($L_X \lesssim 10^{32}$\lumcgs ). Studies of individual systems show that, on average, there is orbital modulation with an overall increase of about a factor of 2 centered around inferior conjunction, often with a hint of a double peaked structure \citep{bog14a,bog14b,gen14,kon12}. However, in most cases the overall count rate is too low to clearly distinguish fine structure to the orbital light curve. Black widows, on the other hand, show a much greater variety in their orbital light curves, with some, like the original black widow PSR B1957+20 \citep{hua12}, having peaks centered around superior conjunction and others around inferior conjunction \citep{gen14}. On average, the redbacks are more luminous than the black widows in X-rays. 

\begin{figure}
\includegraphics[width=55mm, angle=270]{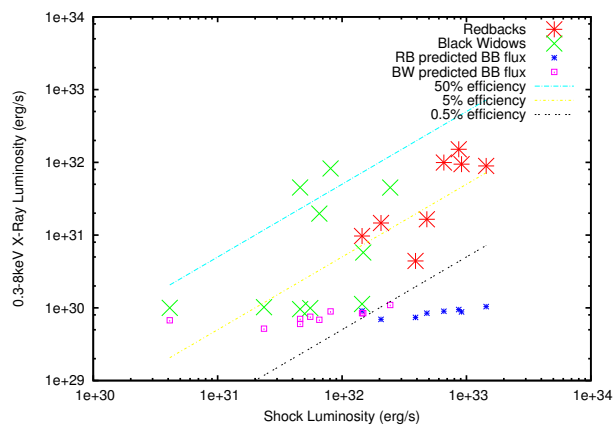}
\caption{Comparison of the X-Ray luminosity from the redback population to that of the black widow population. We define the shock luminosity as  $\dot E \Omega_c$ where $\Omega_c$ is the fraction of the pulsar sky subtended by the companion. We also plot the expected blackbody luminosity for each pulsar, assuming the relationship of \citet{bog15} $L_{bb}=10^{21.28}\dot E^{0.25}$}
\label{shockfig}
\end{figure}

The light curve modeling of \citet{bog11} suggests that the X-ray emission happens very close to the surface of the companion, which suggests that little of the wind that is not directly intercepted by the companion takes part in the X-ray emitting shock. The fraction of the pulsar's sky subtended by the companion, $\Omega_c$, can be calculated from knowledge of the relative masses (requiring knowledge of the orbital inclination angle), the fraction of the Roche lobe the companion fills, and the orbital separation. The inclination angle and Roche lobe filling fraction can be estimated from optical photometric light curves (eg. \citep{bre13}), and when combined with optical radial velocity measurements and the pulsar orbit solution can be used to estimate the masses of the individual components.  In the table, we calculate  $\Omega_c$ from our ``best guess" estimates of neutron star mass, Roche lobe filling factor, and inclination angle using  optical fits where available.  On average, we estimate $\Omega_c\sim 1.3$\% for redbacks and $\Omega_c\sim 0.3$\% for black widows, accounting for much of the relative brightness of the shock emission of redbacks compared to black widows. 

 \begin{table}
 \caption{Redbacks\label{tab1}}
\begin{tabular}{lllllll}

Pulsar & $\log \dot E^a$ & $d^b$& $\Gamma$ &  $\log L_X^c$ & $\Omega_c^d$ & refs\\
 \hline
 J1023+0038 & 34.7 & 1.3 & $1.00^{+0.05}_{-0.08}$ &  32.0  & 1.3\% & (1) \\
 J1227$-$4859 & 35.0 & 1.4 & $1.16^{+0.07}_{-0.08} $ & 31.9 & 1.6\% & (2) \\
 J1628$-$3205 & 34.2 & 1.2 & $1.2^{+0.8}_{-0.7}$ & 31.3 & 1.1\% &  \\
J1723$-$2837 & 34.7  & 0.75 & $1.12^{+0.02}_{-0.02}$ & 32.1 & 2.0\% & (3) \\
J1816+4510 & 34.7 & 4.5 & -- & 31.0 &  0.28\% & (4) (5) \\
J2129$-$0429 & 34.6  & 0.9 & $1.04^{+0.11}_{-0.12}$  & 31.3 & 1.2\% & \\
J2215+5135 & 34.7 & 3.0 & $1.2^{+0.4}_{-0.3}$ & 31.9 & 1.4\% & (6) \\
J2339$-$0533 & 34.4 & 0.4  & $1.09^{+0.40}_{-0.13}$ & 30.6 & 1.6\% & (7) (8) \\
\hline
\end{tabular}

\noindent
\begin{flushleft}
\footnotesize{ a. erg/s b. kpc, from dispersion measure except for J1023+0038 from parallax \citep{del12} and J1816+4510 from optical \cite{kap13} c. erg/s 0.3-8 keV d. estimated percentage of pulsar sky subtended by companion, (1)  \citet{bog11} (2) \citet{bog14a} (3) \citet{bog14b} (4) \citet{sto14} (5)  \citet{kap13} (6)  \citep{gen14} (7)  \citet{rom11}  (8) \citet{ray14}}
\end{flushleft}
\end{table}

We define a ``shock luminosity" as $\dot E \Omega_c$ and plot that vs. the observed 0.3-8~keV X-ray luminosity of redbacks and black widows (Fig.\ref{shockfig}).  We also plot the ``expected" 0.3-8~keV blackbody emission from each pulsar based on  a correlation determined from MSPs 
with good parallax measurements $\log L_{bb}=(0.25\pm0.16)\log \dot E+(21.28\pm 5.36)$ \citep{bog15}. We see that the shock luminosity and X-ray luminosity are correlated, with a typical soft X-ray efficiency relative to the shock luminosity of $\sim 12\%$, albeit with large scatter. We make no estimate of errors in the shock luminosity, being as they are dominated by the very uncertain distances in most cases and a lack of strong constraints from the optical data on inclination and the masses from the optical data in many cases. The redback with the smallest estimated $\Omega_c$ and hence has one of the lowest luminosities is PSR J1816+4510. Optical studies of its companion suggest that it may be a proto-white dwarf which is significantly underfilling its Roche lobe \citep{kap13}. 

Spectrally, the X-ray emission tends to have a constant thermal component, presumably from heated polar caps and consistent with the typical thermal emission from MSPs, and an orbitally variable power-law component.  The fit power-law tends to be very hard with photon spectral index $\Gamma \sim 1$, harder than the typical spectra of pulsar wind nebulae around isolated young pulsars which have $\Gamma \sim 1.5$ in their inner, uncooled regions \citep{kar10}. Below we report on new X-ray observations of two redbacks discovered by the Green Bank Telescope.  

\section{PSR J1628$-$3205}

Discovered in a survey of $Fermi$ sources with the GBT at 820~MHz (Sanpa-Arsa et al. in prep), PSR J1628$-$3205 is a 3.21~ms pulsar in a 5.0~hr orbit around a companion with minimum mass $M_c > 0.16$\ms (assuming $M_{ns}=1.4$\ms ) (Hessels et al. in prep). The pulsar is eclipsed for about 20\% of the orbit. It is modestly energetic with a standard spin-down energy of $\dot E = 1.8\times 10^{34}$~ergs and an estimated distance from the pulse dispersion measure  $d\sim 1.2$~kpc. Optical observations suggest it is Roche lobe filling with minimal heating of the companion \citep{li14}. 

\begin{figure}
 \includegraphics[width=65mm]{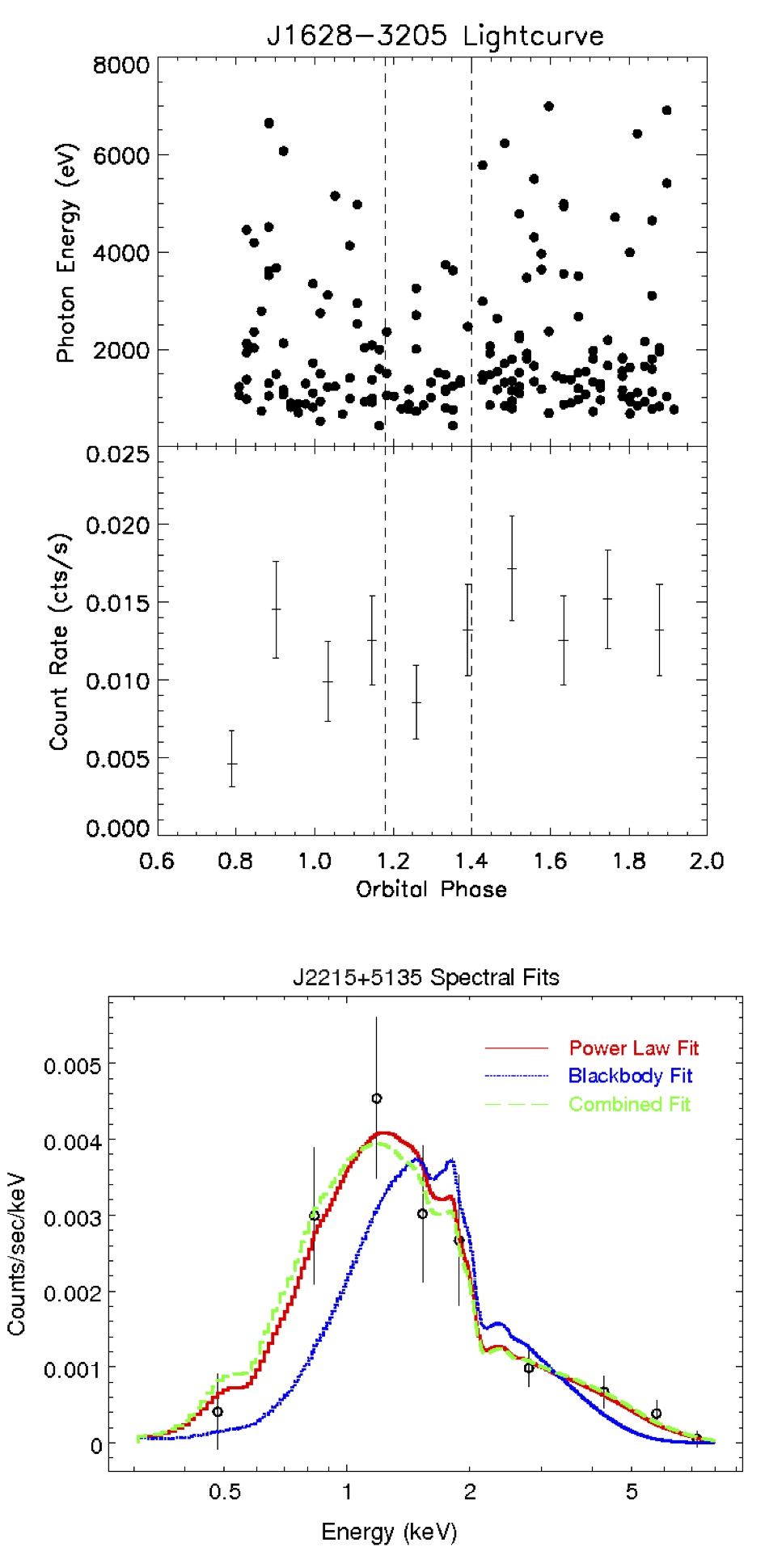}%
 \caption{\label{1628} 20~ks Chandra ACIS-S observation of PSR J1628$-$3205. {\it Top:} Individual photon energies and average count rates as a function of orbital phase. The pulsar superior conjunction is defined as phase 1.25. The dashed lines show roughly the phase range of the radio eclipse. {\it Bottom:} 0.3-8keV spectrum showing absorbed power-law, blackbody, and blackbody+ power-law fits.}
 \end{figure}

We observed PSR J1628$-$3205 for 20~ks (slightly more than one orbit) on 05 May 2012 with the $Chandra$ ACIS-S and detected $\sim 180$ counts.  The counts as a function of orbital phase and energy, plotted in Figure~\ref{1628}, suggest that there may be a dip in the above 2 keV flux near superior conjunction.  The spectrum seems to have a significant  power law component, with a purely blackbody spectrum not giving an acceptable fit. Using the CSTAT statistic of XSPEC (appropriate given  the low number of counts per bin) suggests a pure power law fit provides a somewhat reasonable fit (C-Statistic 24.97 with 22 degrees of freedom), with best fit absorption  
$nH=1.3(0.2-2.5)\times 10^{21} {\rm cm}^{-2}$ and power-law index $\Gamma=1.60(1.23-2.00)$. Using the KS test statistic to determine goodness of fit results in 20\% of realizations having a lower test statistic, suggesting improvements can be made. Since most MSPs have a significant thermal component to their X-ray emission, we next tried an absorbed blackbody plus power-law fit. This resulted in a C-statistic of 21.08 with 20 degrees of freedom, with less than 1\% of KS realizations having a smaller test statistic. The best fit values were $nH=2.2 \times 10^{21} {\rm cm}^{-2}$, $kT=0.20$~keV and $\Gamma=1.14$. The covariance between the blackbody temperature and the power-law index made it difficult to derive reasonable error bars if all parameters were allowed to vary freely, but by constraining the blackbody temperature to vary only between $kT=0.1-0.25$~keV, within which range are the vast majority of MSPs, we find 90\% confidence regions of $nH=(0.3-8.4)\times 10^{21} {\rm cm}^{-2}$ and $\Gamma=(0.5-2.0)$.  The 0.3-8~keV model flux is $F_x=8.8\times 10^{-14} {\rm erg}\, {\rm cm}^{-2}\, {\rm s}^{-1}$ with an unabsorbed flux of $F_x=1.2\times 10^{-13} {\rm erg}\, {\rm cm}^{-2}\, {\rm s}^{-1}$, with roughly 70\% in the power law and 30\% in the blackbody. The fit nH is consistent with the \citet{dri03} Galactic extinction model for a distance of 1.2~kpc. 

\section{PSR J2129$-$0429}

Discovered in a survey of $Fermi$ sources using the GBT at 350~MHz \citep{hes11}, PSR J2129$-$0429 is a 7.61~ms pulsar in a 15.2~hr orbit around a $M_c > 0.37$\ms companion which shows extensive radio eclipses, as much as half the orbit at low frequencies (Hessels et al. in prep). The pulsar has a very high magnetic field for a MSP ($B\sim 1.6\times 10^9$~G), and so still has a high spin down energy $\dot E\sim 3.9\times 10^{34}$ despite its relatively long spin period. The dispersion measure distance is $d\sim 0.9$~kpc.  A variable, bright UV counterpart was evident in the \textit{Swift} UVOT, as was significant X-ray variability from the \textit{Swift} XRT data. Further optical observations suggest the companion is minimally heated and mostly Roche lobe filling and radial velocity measurements suggest a pulsar mass $M_{ns} > 1.7$\ms and a companion mass $M_c\sim 0.5$\ms \citep{bel13}.   These system properties suggest that PSRJ2129$-$0429 is in a relatively early stage in its evolution compared to other redbacks which are more fully spun-up and have typical magnetic fields of a few $10^8$~G. Very large orbital variations are observed through radio timing, and pulsations are dominant in the  $\gamma$-ray emission.

\begin{figure}
\includegraphics[width=75mm]{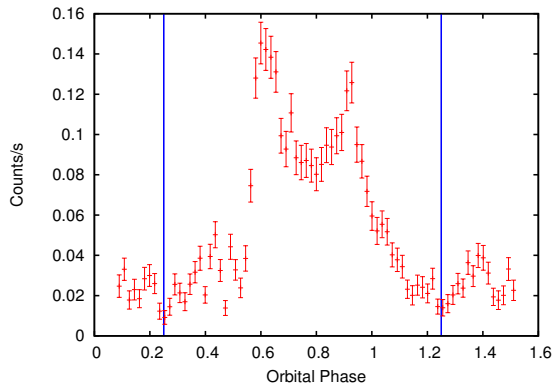}
\caption{\label{2129_lc} $XMM-Newton$ 0.1-10~keV light curve of PSR J2129$-$0429 as a function of orbital phase. The pulsar superior conjunction is indicated by vertical blue lines. }
\end{figure}

\begin{figure}
\includegraphics[width=55mm, angle=270]{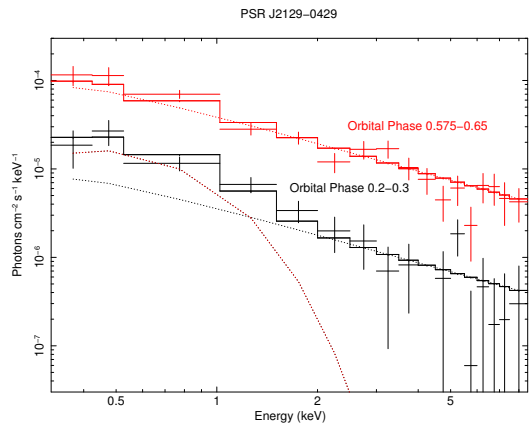}
\caption{\label{2129_spec} Unfolded $XMM-Newton$ PN spectrum of PSR J2129$-$0429 at two orbital phases, keeping the blackbody component fixed.  }
\end{figure}

We observed PSR J2129$-$0429 for 70~ks with $XMM-Newton$. There were no background flares during the observation, meaning we got continuous coverage over slightly more than a complete orbit. The X-ray light curve has very large amplitude variations, with two clear peaks centered on the pulsar's inferior conjunction (Fig.\ref{2129_lc}). We first fit the spectrum with an absorbed blackbody plus power-law, which gave an adequate fit.  The flux is dominated by the power-law component, with an average 0.3-8~keV flux $F_x=2.25\pm 0.05 {\rm erg}\, {\rm cm}^{-2}\, {\rm s}^{-1}$. There is very little absorption ($nH=1.8(0-4.6)\times 10^{20} {\rm cm}^{-2}$) and the thermal component ($kT=0.21(0.16-0.26)$~keV) has a 0.3-8~keV flux $F_{bb}\sim 1.2\times 10^{-14} {\rm erg}\, {\rm cm}^{-2}\, {\rm s}^{-1}$, or about 1/4 of the flux near superior conjunction. The power-law component is very hard ($\Gamma = 1.04(0.92-1.15)$), similar to other redbacks. 
Presuming a constant thermal component throughout the orbit, the difference in the non-thermal flux between the peak at orbital phases 0.575-0.65 and the minimum at phases 0.2-0.3 is about a factor of 11 (Fig. \ref{2129_spec}). There is no evidence of significantly increased absorption. Complete spectral results will be presented in an upcoming paper (Roberts et al. in prep). 

This very remarkable variability suggests that a large fraction of the shock region is blocked by the companion around superior conjunction, suggesting a quite small emission region and a large inclination angle. The two distinct peaks may be a result of doppler boosting and/or relativistic beaming of the synchrotron radiation. The latter would require a strong, well ordered magnetic field. The orbital phases of the peaks, $\sim 0.6$ and $\sim 0.9$, are quite curious. If the shock was wrapped around the companion, then you would expect there to be peaks between phases 0.0-0.5. 
The qualities may suggest a significant role for the magnetic field of the companion. If the companion is tidally locked, like one would expect, then the orbital period of 15.2~hr is the spin period of the companion, which is very rapid. Low mass, rapidly spinning stars can have surface magnetic fields of several hundred to a few thousand Gauss \citep{mor12}. Such potentially large companion fields should not be ignored when investigating the shock emission from redbacks. 

In summary, X-ray emission from the intrabinary shock in redbacks is orbitally dependent, with the increased emission centered on inferior conjunction with potentially a fairly ubiquitous double peaked structure. The emission seems to come from a region that is not much larger than the companion,  is very hard and very efficient, which needs explanation. The previously ignored potential role of the companion's magnetic field in the shock dynamics needs to be considered. 

\bigskip 

\begin{acknowledgments}

Support for this work was provided by the National Aeronautics and Space Administration through Chandra Award Number GO2-13056X issued by the Chandra X-ray Observatory Center, which is operated by the Smithsonian Astrophysical Observatory for and on behalf of the National Aeronautics Space Administration under contract NAS8-03060. This work is based on observations obtained with XMM-Newton, an ESA science mission with instruments and contributions directly funded by ESA Member States and the USA (NASA). 

\end{acknowledgments}

\bigskip 

\begin{thebibliography}{99} 


\bibitem[Archibald et al.(2009)]{arc09} Archibald, A.~M., 
Stairs, I.~H., Ransom, S.~M., et al.\ 2009, Science, 324, 1411 

\bibitem[Arons 
\& Tavani(1993)]{at93} Arons, J., \& Tavani, M.\ 1993, \apj, 403, 249

\bibitem[Bellm et al.(2013)]{bel13} Bellm, E., Djorgovski, 
S.~G., Drake, A.~J., et al.\ 2013, American Astronomical Society Meeting 
Abstracts \#221, 221, \#154.10 

\bibitem[Bogdanov et al.(2011)]{bog11} Bogdanov, S., 
Archibald, A.~M., Hessels, J.~W.~T., et al.\ 2011, \apj, 742, 97

\bibitem[Bogdanov et al.(2014)]{bog14a} Bogdanov, S., Patruno, 
A., Archibald, A.~M., et al.\ 2014, \apj, 789, 40

\bibitem[Bogdanov et al.(2014)]{bog14b} Bogdanov, S., 
Esposito, P., Crawford, F., III, et al.\ 2014, \apj, 781, 6

\bibitem[Bognar et al.(2015)]{bog15} Bognar, K., Roberts, M., 
\& Chatterjee, S.\ 2015, American Astronomical Society Meeting Abstracts, 225, \#346.11

\bibitem[Breton et al.(2013)]{bre13} Breton, R.~P., van 
Kerkwijk, M.~H., Roberts, M.~S.~E., et al.\ 2013, \apj, 769, 108

\bibitem[Deller et al.(2012)]{del12} Deller, A.~T., 
Archibald, A.~M., Brisken, W.~F., et al.\ 2012, ApJL, 756, L25

\bibitem[Drimmel et 
al.(2003)]{dri03} Drimmel, R., Cabrera-Lavers, A., \& L{\'o}pez-Corredoira, M.\ 2003, A\& A, 409, 205

\bibitem[Gentile et al.(2014)]{gen14} Gentile, P.~A., 
Roberts, M.~S.~E., McLaughlin, M.~A., et al.\ 2014, \apj, 783, 69 

\bibitem[Hessels et al.(2011)]{hes11} Hessels, J.~W.~T., 
Roberts, M.~S.~E., McLaughlin, M.~A., et al.\ 2011, American Institute of 
Physics Conference Series, 1357, 40

\bibitem[Huang et al.(2012)]{hua12} Huang, R.~H.~H., Kong, 
A.~K.~H., Takata, J., et al.\ 2012, \apj, 760, 92 

\bibitem[Kaplan et al.(2013)]{kap13} Kaplan, D.~L., Bhalerao, 
V.~B., van Kerkwijk, M.~H., et al.\ 2013, \apj, 765, 158 

\bibitem[Kargaltsev 
\& Pavlov(2010)]{kar10} Kargaltsev, O., \& Pavlov, G.~G.\ 2010, X-ray Astronomy 2009; Present Status, Multi-Wavelength Approach and Future Perspectives, 1248, 25

\bibitem[Kennel \& Coroniti(1984)]{kc84} Kennel, C.~F., \& Coroniti, F.~V.\ 1984, \apj, 283, 694

\bibitem[Kong et al.(2012)]{kon12} Kong, A.~K.~H., Huang, 
R.~H.~H., Cheng, K.~S., et al.\ 2012, ApJL, 747, L3

\bibitem[Li et al.(2014)]{li14} Li, M., Halpern, J.~P., 
\& Thorstensen, J.~R.\ 2014, \apj, 795, 115

\bibitem[Linares(2014)]{lin14} Linares, M.\ 2014, \apj, 795, 
72

\bibitem[Morin(2012)]{mor12} Morin, J.\ 2012, EAS 
Publications Series, 57, 165

\bibitem[Ray et al.(2014)]{ray14} Ray, P.~S., Belfiore, 
A.~M., Saz Parkinson, P., et al.\ 2014, American Astronomical Society 
Meeting Abstracts \#223, 223, \#140.07

\bibitem[Roberts(2011)]{rob11} Roberts, M.~S.~E.\ 2011, 
American Institute of Physics Conference Series, 1357, 127

\bibitem[Romani \& Shaw(2011)]{rom11} Romani, R.~W., \& Shaw, M.~S.\ 2011, ApJL, 743, L26 

\bibitem[Roy et al.(2014)]{roy14} Roy, J., Ray, P.~S., 
Bhattacharyya, B., et al.\ 2014, arXiv:1412.4735

\bibitem[Stappers et al.(2014)]{sta14} Stappers, B.~W., 
Archibald, A.~M., Hessels, J.~W.~T., et al.\ 2014, \apj, 790, 39

\bibitem[Stovall et al.(2014)]{sto14} Stovall, K., Lynch, 
R.~S., Ransom, S.~M., et al.\ 2014, \apj, 791, 67

\bibitem[Tendulkar et al.(2014)]{ten14} Tendulkar, S.~P., 
Yang, C., An, H., et al.\ 2014, \apj, 791, 77 

\bibitem[Thorstensen 
\& Armstrong(2005)]{tho05} Thorstensen, J.~R., \& Armstrong, E.\ 2005, AJ, 130, 759


\end{thebibliography}

\end{document}